\documentclass[12pt]{article}

\def\thefootnote{\fnsymbol{footnote}}
\def\bea{\begin{eqnarray}}
\def\eea{\end{eqnarray}}
\def\beq{\begin{equation}}
\def\eeq{\end{equation}}
\def\Tr{{\rm Tr}}

\def\ux{$U(1)_X$}
\def\qx{q_X}
\def\gx{g_X}
\def\Dx{D_X}

\def\M{\bar{M}}

\def\bv{\bar{v}}
\def\bj{\bar{\jmath}}
\def\bbi{\bar{\imath}}
\def\[{\left [}
\def\]{\right ]}
\def\<{\left <}
\def\>{\right >}
\def\({\left (}
\def\){\right )}

\def\dsq{|\delta|^2}
\def\dfr{|\delta|^4}
\def\tG{{\tilde G}}
\def\tu{{\tilde u}}
\newcommand{\ben}{\begin{enumerate}}
\newcommand{\een}{\end{enumerate}}

\newcommand{\mtxt}[1]{\mathop{\hbox{{\small #1}}}\nolimits}
\newcommand{\ttxt}[1]{\mathop{\hbox{{\tiny #1}}}\nolimits}

\newcommand{\beqa}{\begin{eqnarray}}
\newcommand{\eeqa}{\end{eqnarray}}

\newcommand{\p}{{\partial}}

\newcommand{\vev}[1]{{\langle #1 \rangle}}
\newcommand{\bigvev}[1]{{\left\langle #1 \right\rangle}}

\newcommand{\ordnt}[1]{{{\cal O}(#1)}}
\newcommand{\myref}[1]{(\ref{#1})}
\begin{document}

\begin{titlepage} 
\begin{center}
            \hfill    LBNL-44856 \\
            \hfill    UCB-PTH-00/02 \\
            \hfill hep-ph/0001219\\
	\hfill \today\\

\vspace{18pt}
{\bf \Large A D-moduli problem?}

\vspace{18pt}
 Mary K. Gaillard\footnote{E-Mail: {\tt MKGaillard@lbl.gov}}
{\em and} Joel Giedt\footnote{E-Mail: {\tt JTGiedt@lbl.gov}}

\vspace{18pt}

{\em Department of Physics, University of California, \\
and Theoretical Physics Group, 50A-5101, \\
Lawrence Berkeley National Laboratory, Berkeley, 
CA 94720 USA.}\footnote{This work was supported in part by the
Director, Office of Energy Research, Office of High Energy and Nuclear
Physics, Division of High Energy Physics of the U.S. Department of
Energy under Contract DE-AC03-76SF00098 and in part by the National
Science Foundation under grant PHY-95-14797.}\\[.1in]

\vspace{18pt}

\end{center}

\begin{abstract}

We point out a
generic problem in string-inspired supergravity
models with an anomalous $U(1)_X$.
A large number of matter multiplets charged under $U(1)_X$
remain massless above the supersymmetry-breaking
scale because of degeneracy of vacua solving the
D-flatness conditions.
A toy model is analyzed as an illustration
of the mechanism; we find the surprising
result that many scalars remain massless
after supersymmetry-breaking in a hidden
sector.
\end{abstract}
\end{titlepage}

\newpage
\renewcommand{\thepage}{\roman{page}}
\setcounter{page}{2}
\mbox{ }

\vskip 1in

\begin{center}
{\bf Disclaimer}
\end{center}

\vskip .2in

\begin{scriptsize}
\begin{quotation}
This document was prepared as an account of work sponsored by the United
States Government. Neither the United States Government nor any agency
thereof, nor The Regents of the University of California, nor any of their
employees, makes any warranty, express or implied, or assumes any legal
liability or responsibility for the accuracy, completeness, or usefulness
of any information, apparatus, product, or process disclosed, or represents
that its use would not infringe privately owned rights. Reference herein
to any specific commercial products process, or service by its trade name,
trademark, manufacturer, or otherwise, does not necessarily constitute or
imply its endorsement, recommendation, or favoring by the United States
Government or any agency thereof, or The Regents of the University of
California. The views and opinions of authors expressed herein do not
necessarily state or reflect those of the United States Government or any
agency thereof of The Regents of the University of California and shall
not be used for advertising or product endorsement purposes.
\end{quotation}
\end{scriptsize}

\vskip 2in

\begin{center}
\begin{small}
{\it Lawrence Berkeley Laboratory is an equal opportunity employer.}
\end{small}
\end{center}

\newpage
\renewcommand{\thepage}{\arabic{page}}
\setcounter{page}{1}
\def\thefootnote{\arabic{footnote}}
\setcounter{footnote}{0}

In this Letter we consider a simple toy model which illustrates a
generic problem in string-inspired supergravity
models with an anomalous $U(1)_X$.
A large number of matter multiplets charged under $U(1)_X$
remain massless above the supersymmetry-breaking
scale because of degeneracy of vacua solving the
D-flatness conditions.  We refer to these
multiplets as ``D-moduli.''  For example, in the model described
in Section 4.2 of~\cite{font:90a}, that we will refer to here as the
FIQS model, there are 26 massless chiral multiplets
associated with this degeneracy. In the toy model considered here,
we find that the degeneracy is partially broken
by the introduction of a generic supersymmetry-breaking
term in such a way that the overall vacuum energy vanishes.
However many scalar fields remain massless even
after supersymmetry breaking. Specifically, in our toy model with 
the convention\footnote{Our charge normalization is such that $\Tr T^2_a =
{1\over2}$ if $T_a$ is a generator of $SU(N)$.} that 
$\Tr Q_X >0$, where $Q_X$ 
is the generator of $U(1)_X$, the only remaining
flat directions are those for which a linear combination of
fields with the lowest value $q_0$ of the \ux\, charge acquires
a {\it vev}.  The corresponding linear combination of
chiral multiplets forms a massive vector multiplet with the
\ux\, gauge fields, while the orthogonal combinations are massless.
In the FIQS model, for example,
the lowest \ux\, charge is $-8$ with a 15-fold degeneracy, so 14
complex scalars (as well as 26 chiral fermions) would be massless
if no other symmetries were broken.
The scalar fields with $\qx>q_0$ would acquire soft masses 
of the same order as generically expected for squarks and sleptons.
However in string-derived models such as this one, the D-moduli are 
charged under other $U(1)$'s, which partially lifts the degeneracy 
of the vacuum before supersymmetry breaking.

We first consider our toy model in the context of standard
supergravity with scalars and their superpartners in chiral
multiplets. Then we appeal more explicitly to the linear multiplet
formulation of gaugino condensation~\cite{bgw} as the mechanism of
supersymmetry breaking, which we refer to as the BGW model.  Neither
of our models is realistic. For example, we neglect the dependence of
the matter K\"ahler metric on T-moduli ({\it i.e.}  breathing
modes). While including this would considerably complicate the
analysis, it seems unlikely to provide a mechanism (at least at tree
level) for lifting the degeneracy of the vacuum.  We also neglect
additional, nonanomalous $U(1)$ couplings of the fields with large
$vev$'s; as discussed below, we do not expect them to lift the
vacuum degeneracy completely.  We illustrate this point using the FIQS
model, which itself should probably be considered a toy model for reasons
discussed below. Finally we suggest a possible mechanism for lifting
the remaining degeneracy after supersymmetry breaking, and comment on
implications for cosmology.

For our toy chiral supergravity model, we assume a K\"ahler potential
\beq
K = \sum_i |A_i|^2 + \sum_i |B_i|^2 + \sum_i |\Phi_i|^2
+ K(M,\M),
\label{2.1}
\eeq
a superpotential
\beq
W = \lambda_{ijk} A_i B_j \Phi_k + W(M),
\label{2.2}
\eeq
and a gauge group $G_{\ttxt{gauge}} = SU(3)_c \times U(1)_X$.  
We denote $U(1)_X$ charges by:
\beq
Q_X A_i = n_i A_i, \quad Q_X B_i = p_i B_i, \quad
Q_X \Phi_i = q_i \Phi_i,\quad Q_X M_a = 0.
\label{2.3}
\eeq
We assume $q_i \not=0$ and $|q_i| \sim \ordnt{1}$.  We further
assume that the $q_i$ are such that $<\Dx> = 0$ has a solution with
$<A_i> = <B_i> = 0$; in other words we assume that there are flat
directions that allow supersymmetry to remain unbroken in the absence 
of the nonperturbatively induced superpotential $W(M)$, as was found in the
string-derived models studied in~\cite{font:90a}. Scalar
components of the gauge-charged superfields are given by $a_i =
A_i|$, $b_i = B_i|$, $\phi_i = \Phi_i|$.
We take $A_i$ and $B_i$ to be charged under
$SU(3)_c$, while $\Phi_i$ and $M_a$ are $SU(3)$ singlets.  
{\it E.g.}, $A_i$ is a
${\bf 3}$ while $B_i$ is a ${\bf \bar 3}$.  Couplings such
as \myref{2.2} occur in semi-realistic
heterotic orbifold models \cite{font:90a}.
When $\vev{\phi_i} \not= 0$, generally required by D-flatness
because of the Fayet-Illiopoulos (FI) term associated with
\cite{dine:87} the anomalous $U(1)_X$, some
color triplets acquire large masses.  
We demand that $SU(3)_c$ remain unbroken at all
scales.  Therefore
\beq \vev{a_i} = \vev{b_i} = 0 . \label{2.4} \eeq
We interpret the gauge singlet superfields $M_a$ as
moduli; the supersymmetry-breaking superpotential
$W(M)$ in \myref{2.2}:
\beq \vev{W(M)|} \equiv \delta, \label{2.5} \eeq
is assumed to be generated by nonperturbative dynamics. Cancellation of
the $U(1)_X$ anomaly by the Green-Schwarz (GS) mechanism induces an FI 
term $\xi$ in the $U(1)_X$ D-term \cite{dine:87}:
\beq D_X = \sum_i q_i |\phi_i|^2 + \sum_i n_i |a_i|^2
+ \sum_i p_i |b_i|^2 + \xi, \quad \xi = {\gx^2\over24\pi^2}\Tr Q^3_X  . \label{2.6} \eeq
Motivated by semi-realistic models of string-derived
effective supergravity with dynamical supersymmetry-breaking,
we assume (in units where $m_P = 1/\sqrt{8\pi G} = 1$):
\beq
|\delta|^2 \ll |\xi| \ll 1 .
\label{2.7}
\eeq
In addition to $D_X$, we also have the $SU(3)_c$ D-term
$D_c^{(r)}$, $r=1,\ldots, 8$, which does not play a
role in the following analysis.  Let capital
indices $I,J$, {\it etc.}, refer collectively to fields $a_i, b_i,
\phi_i$.  Then the scalar potential of the toy model is
given by:
\bea
V &=& {g_X^2 \over 2} D_X^2 + {g_c^2 \over 2} D_c^{(r)} D^c_{(r)}
+ e^K \bigg[ \delta^{I \bar J} (W_I + W K_I)(\bar W_{\bar J} +
\bar W K_{\bar J} ) \nonumber \\ & & 
+ K^{a \bar b} (W_a + W K_a)(\bar W_{\bar b} +
\bar W K_{\bar b})- 3 W \bar W \bigg].
\label{2.8}
\eea
We parameterize the vacuum value of the moduli sector F-term
as 
\beq \left<K^{a \bar b} (W_a + W K_a)(\bar W_{\bar b} +
\bar W K_{\bar b})\right> = \alpha |\delta|^2, \quad \alpha\sim 1.\eeq
The requirement of a vanishing cosmological constant gives
\bea <V> &=& {\gx^2\over2}\<\Dx^2\> + e^{<K>}|\delta|^2\(v^2 + \alpha - 3\)
= 0, \nonumber \\ \<\Dx\> &=& \sum_i q_i |v_i|^2 + \xi,\label{vac}\eea
where $v_i = \vev{\phi_i}$ and $v = \sqrt{\sum_i |v_i|^2}$.
Since $V$ is gauge neutral, $\vev{\p V / \p a_i}$
and $\vev{\p V / \p b_i}$ are $SU(3)$-charged and vanish when
\myref{2.4} holds. The minimization condition for $\phi^i$ gives
\beq <V_i>\equiv \bigvev{{\p V \over \p \phi_i}} = 0 =
\bar v_i \left[ g_X^2 q_i\<\Dx\> + |\delta|^2 e^{<K>} ( v^2 + \alpha - 2) 
\right] .\label{3.1}\eeq
This implies that $v_i\ne0$ for only one value of $q_i\equiv -q$, which,
as we shall see, must be negative under our assumption \myref{2.7}.
Note that \myref{3.1} require $\<\Dx\>\sim\dsq\ll <|W|>$, so that supersymmetry
breaking is dominated by the moduli sector under our assumptions.  
(\ref{3.1}) and (\ref{vac}) together imply 
\beq v^2 + \alpha - 3 = -O(\dsq),\quad v^2 + \alpha - 2 \approx 1,\eeq
so that $\<\Dx\>>0$.  Next we consider the spectrum of $\Phi^i$.
The fermion mass matrix takes the form 
\beq (M^2_f)^i_j = v^i\bv^{\bj}\[2\gx^2q^2 + e^{<K>}\dsq(v^2 + \alpha)\],
\label{fmass}\eeq
and the scalar mass matrix (in Landau gauge) takes the form
\bea M^2_s &=& \pmatrix{M^2 & N^2\cr (N^{\dag})^2 & (M^{\dag})^2},
\nonumber \\ 
(M^2)^i_j &=& (M^2_f)^i_j - v^i\bv^{\bj}\gx^2q^2 + \delta^i_j\[
\gx^2q_i\<\Dx\> + e^{<K>}\dsq\(v^2 + \alpha - 2\)\]
\nonumber \\  
(N^2)^{\bbi}_j &=& \bv^{\bbi}\bv^{\bj}\[\gx^2q^2 
+ e^{<K>}\dsq(v^2 + \alpha - 1)\].\label{smass}\eea
In the absence of supersymmetry breaking, $\<\Dx\> = \delta = 0$, the superfield
\beq \Pi = {1\over v}\sum_i\bv^i\Phi^i\eeq
is eaten by the \ux\, gauge supermultiplet to form a massive vector multiplet.
The orthogonal combinations
\beq D_\alpha = \sum_i c^i_\alpha\Phi^i, \quad \sum_iv^ic^i_\alpha = 0,
\quad \sum_i \bar c^i_\alpha c^i_\beta = \delta_{\alpha \beta},\eeq
are the massless D-moduli.  When $\delta\ne0$, the moduli $M_a$ mix with $\Pi$; in
other words the ``Goldstone'' chiral multiplets associated with supersymmetry
breaking and with \ux\, breaking mix at order $\dsq/\gx^2q^2v^2$.  There is
no mixing between the D-moduli and the M-moduli, so the D-moduli masses
can be read directly from \myref{fmass} and \myref{smass} by setting $v_i=0$;
one obtains $M^2_{Df} = 0,$ and, using the vacuum conditions \myref{vac}
and \myref{3.1}, for the scalars $d_\alpha = D_\alpha|$:
\beq m^2_\alpha = (q_\alpha + q)\gx^2<\Dx> = {(q_\alpha + q)\over q}
\[m^2_{\tG} + O(\dfr)\],\eeq
where 
\beq m_{\tG} = e^{<K>/2}\delta \eeq
is the gravitino mass.  From (17) it is clear that $m_\alpha^2 \geq 0$
if and only if $q = - \min (q_1, \ldots ,q_{N_\Phi} )$.
The $d_\alpha $ that are linear combinations of $\phi^i$ with
$q_i = -q$ remain massless, while the others acquire masses of order of the gravitino
mass.

We now turn to a more specific model for supersymmetry breaking {\it via}
gaugino condensation, as realized in the linear multiplet formulation for
the dilaton~\cite{bgw}.  Our model is an approximation to the BGW model, 
in that we neglect the moduli-dependence of the K\"ahler metric for the 
gauge-charged matter fields that we consider.  With this approximation,
the scalar potential takes the form\footnote{The full scalar potential
for the BGW model is given in~\cite{glm}.}
\bea V &=& {1\over2}\gx^2\Dx^2 
+ \sum_I\left|(W_I + K_I W)e^{K/2} + \beta uK_I\right|^2
\nonumber \\ & &
+ f(\ell)\left|u(1 + b_a) - 4\ell W e^{K/2}\right|^2
- {3\over16}\left|u b_a - 4 W e^{K/2}\right|^2, \label{linpot}\eea
where $u = e^{K/2}\tu(\ell,t)$ is the gaugino condensate which is 
determined by the equations of motion~\cite{bgw} as a function
of the dilaton $\ell$ and the T-moduli $t$ (treated here as 
constants\footnote{When the dilaton is dynamical, $\gx \to \gx(\ell)$ and
$\gx^2\xi \to 2\ell\xi$. (When nonperturbative string effects are
neglected, $\gx^2(\ell) = 2\ell$ at the string scale.)}), 
$W = W(\phi)$, $b_a$ is the $\beta$-function 
coefficient for the condensing gauge group, and the function $f(\ell)$
depends on the K\"ahler potential for the dilaton.  
The terms in \myref{linpot}
are, in a one-to-one correspondence, the counterparts in this model of the
terms in \myref{2.8}.  The vacuum conditions are, assuming $<W(\phi)>=
<W_I>=0$,
\bea <V> &=& \left<{1\over2}\gx^2\Dx^2 
+ f(\ell)|u|^2(1 + b_a)^2 - {3\over16}|u|^2 b_a^2 + \beta^2|u|^2v^2\right> 
\nonumber \\ &\equiv& \left<{1\over2}\gx^2\Dx^2\right> + \hat{V} =0, 
\nonumber \\ 
<V_i> &=& \bv^{\bbi}\(q_i\left<\gx^2\Dx\right> + \hat{V} + \beta^2<|u|^2>\).
\label{linvac}\eea
In the spirit of the previous section, we assume \myref{2.7} with
$<|u|>\sim|\delta|$, so that $\Dx\sim\dsq,\;\hat{V} \sim \dfr$.  As in the
previous example, one chiral supermultiplet with $q_i = -q$, the lowest
\ux\, charge, forms a massive gauge supermultiplet with the \ux\, gauge
superfield, while the remaining chiral superfields  
have massless fermions and scalar masses now given by
\beq m^2_\alpha = (q_\alpha + q)\gx^2<\Dx> = {(q_\alpha + q)\over q}
\[\({4\beta\over b_a}\)^2m^2_{\tG} + O(\dfr)\].\eeq
When the GS term that cancels the T-duality anomaly is 
included, the parameter $\beta$ is given by 
\beq \beta = (p_\alpha - b_a)/4,\label{beta}\eeq
where $p_\alpha$ measures the coupling of the fields $\phi^i$ to the GS 
term.  Here the situation is the same as for squarks and sleptons;
if $p_\alpha = 0$, $m_\alpha \approx m_{\tG}$, while if the D-moduli 
couple to the GS term with the same strength $b$ as the T-moduli, 
$m_\alpha \approx m_t/2 \approx |b/b_a - 1|m_{\tG}$.  
In the BGW model with 
$b = b_{E_8} \approx 10b_a$ we get $m_\alpha \approx 10m_{\tG}$.
However, in the presence of Wilson lines that break the gauge group to a
phenomenologically viable one, in general~\cite{lust} $b<b_{E_8}$.  In 
particular, in the FIQS model discussed below, with an $SO(10)$
condensing group, $b = b_a$, which implies that the moduli masses are 
much smaller than the gravitino mass, so the FIQS model is not viable in
the context of the BGW supersymmetry breaking scenario.


As mentioned above, in more realistic models the D-moduli are charged
under additional, nonanomalous gauge groups $U(1)_a$.  Assuming affine
level one, so that the gauge couplings are all equal at the string
scale:\footnote{Strictly speaking we should integrate out the heavy
modes at the scale $v^i \sim .1$ and run the couplings their values to
the condensation scale $<u^{1\over3}>\sim 10^{-4}$; we neglect such
renormalization effects here.} $g_a=\gx\equiv g$, 
the potential \myref{2.8} or \myref{linpot} takes the form
\beq V = {g^2 \over 2}\sum_aD_a^2 + \hat V.\label{fiqspot}\eeq
where we have set to zero D-terms corresponding to nonabelian gauge
groups such as $SU(3)$ in our toy model.  The minimization conditions
take the form
\beq 0 = <V_i> = g^2 \sum_a<D_a> q_i^a \bar v_i + <\hat V_i> .
 \label{fiqsmin}\eeq 
In the models we are considering, we may write 
\beq <\hat V_i> = \bar v_i f, \label{simp} \eeq
where $f$ is some function of the $v^i$ and the
moduli $ vevs$'s.  Then for any $i$ such that $v^i \not= 0$,
\myref{fiqsmin} implies:
\beq 0 = g^2\sum_a D_a q_i^a + f =  g^2\sum_aq_i^a\sum_j|v^j|^2
+ g^2q^X_i\xi + f. \label{mini2} \eeq
We are interested in models with F-flat and D-flat directions, {\it i.e.}
in which the set of equations 
\beq D_a = 0 \label{d=0}\eeq
 has a solution along some F-flat
direction.  For example in the FIQS model, with $a=1,\ldots 8,$ 
(a continuous degeneracy of)
solutions exist with nonvanishing $vev$'s for a set of 27 complex scalar
fields\footnote{Our notation differs slightly from that 
of~\cite{font:90a}.}
\beq \Phi^i \equiv S_\alpha^i,\;Y_A^i,\quad \alpha = 1,\cdots 5,\quad 
A=1,\cdots 4,\quad i=1,\cdots 3,\eeq
which are charged only under the $U(1)_a$, with the charges
$q_a$ independent of the index $i$ and
\beq \qx^S = -8, \quad \qx^Y = 4, \quad q_a^{Y_1} = q_a^{Y_2},\quad
q_a^{Y_3} = q_a^{Y_4}.\eeq
The set of equations \myref{d=0} have solutions that break 6 of the 8
$U(1)$'s, including \ux, leaving unbroken the weak hypercharge of the
Standard Model and one additional $U(1)$; in the effective low energy 
theory at scales $\mu\ll\xi$ there are no supermultiplets that carry
both this latter $U(1)$ charge and Standard Model gauge 
charges~\cite{font:90a}.  Since 6 $U(1)$'s are broken, 6 of the 
supermultiplets in (28) are eaten by massive vector multiplets, and we
are left with 21 D-moduli supermultiplets, instead of the 26 we would have
in the absence of additional $U(1)_a$ charges for these fields.

Now consider the effect of the supersymmetry breaking term $f$ in
\myref{fiqsmin}.  The solution to \myref{d=0} for $a\ne X$
requires that at least one field $Y$ have
a nonvanishing vacuum value.  Therefore $D_{a\ne X}=0$ is not a solution 
to \myref{fiqsmin} since the previous analysis without the additional
$U(1)_a$ requires in this case that only $<S>\ne 0$ when $f\ne0$.  Hence 
we are led to solve\footnote{Again we are oversimplifying;  once 
supersymmetry is broken there is no reason to assume that the F-terms
involving D-moduli couplings in the superpotential remain zero.} the set
\myref{mini2} of coupled equations for the $|v^i|^2$.
We have analyzed these equations using the math package Maple and the
$U(1)_a$ charge assignments of the FIQS model, and find that the minimum
corresponds to 
\bea && Y_3^i = Y_4^i = 0, \quad \sum_i|S^i_\alpha|^2 = f_\alpha(\xi,g^2,f), \nonumber \\
&& \sum_i\(|Y^i_1|^2 + |Y^i_2|^2\) = f_Y(\xi,g^2,f),\label{constr}\eea
with an additional constraint of the form $f = f(g^2,\xi)$ to assure
vanishing of the cosmological constant.
Now $Y_1^i$ and $Y_2^i$ correspond to 12 real fields constrained by one 
equation to give 11 moduli, and each of the 5 choices of $\alpha$
in $S_\alpha^i$ correspond to 6 real fields subject to one constraint
giving 5 moduli each.  In this model there are 6 $U(1)$'s that
get broken, so 6 moduli are eaten, leaving a total of 
$$5 \times 5 + 11 - 6 = 30$$ D-moduli.  Note that while there are fewer
``light'' D-moduli ($m\sim m_{\tG}$) than in the toy model with the same
\ux\, charges but no additional $U(1)$'s, there are actually more massless
scalars (30 instead of 28) after supersymmetry breaking.

We remark that the first condition in \myref{constr}
(which corresponds to $Y_3^{1i} =
Y_3^{2i}= 0$ in the notation of~\cite{font:90a})
has as a consequence that all of the down-type
quarks are massless at tree level, and their masses must
be generated by radiative corrections.  Leaving aside the moduli problem
alluded to above, this could be a phenomenological
improvement of the model as compared with the solution of \myref{d=0}
in the absence of supersymmetry breaking. In that case there are both
up- and down-type quark masses at tree level, but 
(unless unmotivated mixing is introduced in the
K\"ahler potential) the CKM matrix is unrealistic: the heaviest up quark
is not in the same $SU(2)_L$ gauge multiplet as the heaviest down quark.
It has recently been shown that all down-type quark
masses can be generated entirely from radiative
corrections, subject to certain conditions on the
high energy theory and supersymmetry
breaking scenario~\cite{borzu:99a}.
Whether or not viable quark masses can be gotten
by this mechanism in the FIQS model is under
investigation.

In the generic gaugino condensation model of~\cite{bgw}, supersymmetry
breaking arises from the Vene\-ziano-Yan\-kiel\-o\-wicz part of the
superfield Lagrangian:
\beq {\cal L}_{\mtxt{VY}} = \int d^4\theta{E \over 8R} U
	\left[ b' \ln ( e^{-K/2}U) + \sum_\alpha b_\alpha
	\ln \Pi_\alpha \right] + \mtxt{h.c.}.\eeq
The values of $b'$ and $b_\alpha$ are determined
by anomaly matching and are related to the $\beta$-function coefficient by
\beq b_a = b' + \sum_\alpha b_\alpha. \eeq
The (weight two) chiral field $U$ is the gaugino condensate superfield:
$U| = u$, while the (weight zero) chiral fields $\Pi$ are matter 
condensates.  Condensation occurs
provided there is also a superpotential for the matter condensates:
\beq W(\Pi,T) = \sum_\alpha c_\alpha(T) \Pi_\alpha,\eeq
where the moduli-dependence of the coefficient assures modular invariance.
In~\cite{bgw} it was assumed that $\Pi$ is a composite operator containing
fields charged only under the condensing gauge group.  However in many
models -- such as the FIQS model with a hidden sector $SO(10)$ and matter
in 16's -- there is no operator that can be constructed from these fields
alone that is invariant under the $U(1)_a$, and the coefficient
$c_\alpha$ must depend on the $\Phi^i$.  
It is possible that these additional couplings of the 
D-moduli are sufficient to lift the remaining degeneracy of the vacuum
--  and they may also give $O(m_{\tG})$ masses to the D-moduli fermions.
This is because \myref{simp} would no longer hold, so that we do
   not get the single condition \myref{mini2}, but rather
   several independent conditions from \myref{fiqsmin}.
An analysis of this case requires a careful treatment of renormalization
effects.  However, it appears
likely that any masses generated by these additional couplings will be
governed by the supersymmetry breaking scale.

A large number of light scalar fields is problematic for cosmology in
realistic models.  The D-moduli have no gauge couplings in the
effective low energy theory since they are charged only under 
the $U(1)$'s that are broken near the
string scale.  Therefore, unless they have unsuppressed superpotential
couplings to relatively light particles, they are subject to the same
constraints as, {\it e.g.,} the
T-moduli~\cite{cough:83}--\cite{gonch:84}.  The problem is somewhat
alleviated if they couple to the Green-Schwarz term with $p_\alpha = b$
in~\myref{beta} and $b\gg b_a$.  In this case, like the T-moduli in
the BGW model, their masses can exceed the gravitino mass by an order
of magnitude.  
In $Z_3$ and $Z_7$ compactifications, with no T-moduli-dependent
string threshold corrections~\cite{tom}, $b\ge b_a$ where the inequality is saturated
if there are no twisted sector fields that are charged under the gauge
group ${\cal G}_a$.  This is the case, in particular for the FIQS model 
that we used above to illustrate the case of many $U(1)$'s.  As we noted
previously, this is not a viable option in the BGW context, since it gives
unacceptably small moduli masses.  In addition, models with $b\gg b_a$
alleviate~\cite{bgw} problems associated with dilaton cosmology.

It is plausible that the fermions can be
sufficiently diluted by inflation to be harmless. The decay and
annihilation of the particles of the Minimal Supersymmetric Standard
Model suppresses the energy density of a decoupled massless fermion
relative to a neutrino only by a factor of about 20, so 20 such
fermions would contribute the equivalent of one neutrino species to the
energy density during Nucleosynthesis if there is no other suppression
mechanism.  Fermions with order $TeV$ masses would vastly overclose
the universe unless they are inflated away or are sufficiently short
lived.  With only gravitational strength couplings, dimensional analysis
suggests decay rates $\Gamma_\alpha\sim m_\alpha^3/m_P^2$, which were
shown~\cite{ccqr} to be marginally acceptable for the T-moduli and
dilaton fermions.  Any remnant of the broken $U(1)$'s in the tree-level 
couplings of the D-moduli would tend to suppress the decay rate. For 
example $U(1)$ invariant couplings would give rates
$\Gamma_\alpha\sim m_\alpha^5/m_P^4$ that are unacceptably small without
sufficient dilution.
  An estimate of the D-moduli fermion lifetimes requires a detailed
understanding of the effective theory below the scale where the
$U(1)_a$ are broken.

The point that we wish to emphasize here is that there are, generically, 
many more light moduli than have been previously considered, which may
imply much stronger constraints on their masses and/or couplings.
A full analysis of the D-moduli spectrum in the context of the BGW model
for supersymmetry breaking, including dynamical T-moduli and dilaton, 
will be given elsewhere.

\end{document}